\documentclass[sigconf, nonacm]{acmart}

\usepackage{url}
\usepackage{subcaption}
\usepackage{enumitem}
\usepackage{graphicx}
\usepackage{caption}
\usepackage{titlesec}
\usepackage{multirow}

\usepackage[normalem]{ulem}

\setlist{
  listparindent=\parindent,
  parsep=0pt,
}

\begin{document}

\date{}

\title{ELMo-Tune-V2: LLM-Assisted Full-Cycle Auto-Tuning to Optimize LSM-Based Key-Value Stores}

\author{
{\rm Viraj Thakkar$^1$, Qi Lin$^1$, Kenanya Keandra Adriel Prasetyo$^2$, Raden Haryosatyo Wisjnunandono$^3$, Achmad Imam Kistijantoro$^3$, Reza Fuad Rachmadi$^2$, Zhichao Cao$^1$}\\
$^1$Arizona State University, $^2$Institut Teknologi Bandung, $^3$Sepuluh Nopember Institute of Technology
\and
}

\begin{abstract}

Log-Structured Merge-tree-based Key-Value Store (LSM-KVS) is a foundational storage engine serving diverse modern workloads, systems, and applications. To suit varying use cases, LSM-KVS allows a vast configuration space that controls core parameters like compaction, flush, and cache sizes, each consuming a shared pool of CPU, Memory, and Storage resources. Navigating the LSM-KVS configuration space necessitates knowledge of the impact of each configuration on the expected workload and underlying hardware. Beyond expensive and time-intensive human-expert-based tuning, existing LSM-KVS tuning solutions focus on tuning with specific workload expectations while limited to a narrow subset of parameters.

This paper introduces ELMo-Tune-V2, a framework that integrates Large Language Models (LLMs) at its foundation to demonstrate the potential of applying modern LLMs in data system optimization problems. ELMo-Tune-V2 leverages the contextual reasoning, cross-domain, and generative capabilities of LLMs to perform 1) self-navigated characterization and modeling of LSM-KVS workloads, 2) automatic tuning across a broad parameter space using cross-domain knowledge, and 3) real-time dynamic configuration adjustments for LSM-KVS. ELMo-Tune-V2 integrates three innovations: LLM-based workload synthesis for adaptive benchmark generation, feedback-driven iterative fine-tuning for configuration refinement, and real-time tuning to handle evolving workloads. Through detailed evaluation using RocksDB under several real-world applications across diverse scenarios, ELMo-Tune-V2 achieves performance improvements up to $\sim14X$ our YCSB benchmarks compared against default RocksDB configurations, and our end-to-end tests with upper-level applications, NebulaGraph and Kvrocks, demonstrate performance gains of 34\% and 26\%, respectively
\end{abstract}

\maketitle

\section{Introduction}

The modern Large Language Model (LLM) has emerged as a transformational tool in Artificial Intelligence (AI), showcasing unparalleled capabilities in synthesizing structured and unstructured information, performing contextual reasoning, and demonstrating situational adaptability \cite{liu_understanding_2024}. Its advent has revolutionized fields closely related to AI like natural language processing \cite{minaee_large_2024}, decision support \cite{eigner2024determinants}, and generative modeling \cite{openai_gpt-4_2024}. However, the potential of LLM in resolving and optimizing data system challenges, particularly in domains that demand analysis of structured metrics and system expertise, remains under-explored. This paper investigates the prospect of LLM in optimizing Log-Structured Merge-tree-based Key-Value Stores (LSM-KVS), a cornerstone of modern data storage systems.

LSM-KVS systems, such as RocksDB \cite{rocksdb}, LevelDB \cite{leveldb}, and HBase \cite{hbase}, underpin critical upper-layer applications ranging from Meta's ZippyDB \cite{masti2021we} to graph databases like Nebula \cite{wu2022nebula}. The log-structured design enables high write throughput and extensive tunability, making them ideal for diverse workloads. However, this flexibility introduces significant complexity: LSM-KVS comprises numerous interdependent components (e.g., compaction, flush, and foreground I/Os), with configuration spaces spanning more than one hundred option parameters as in popular LSM-KVS like RocksDB and HBase. Further, achieving optimal performance requires addressing the intricate interplay between hardware (e.g., SSD vs. HDD), resource limits (e.g., 2GiB vs 4GiB DRAM), and workload (e.g., read-heavy vs. write-heavy) dynamics. 

Companies employ domain experts who utilize a multi-stage process to optimize LSM-KVS. This end-to-end optimization process begins with a comprehensive understanding of the workload through characterizing and analyzing access patterns, read/write ratios, and data distributions \cite{cao2020characterizing, kim_robust_2020, luo_breaking_2020, dayan_monkey_2017, Dostoevsky, ma_query-based_2018}. Concurrently, experts must understand the underlying hardware, system architecture, and deployment environments, including storage media types and server configurations \cite{dong2023disaggregating, dong2021evolution, wu2018data, wu2019zonealloy}. Building upon these insights, they analyze behavioral patterns (e.g., key-value sizes, query intensiveness, and access distributions) and internal correlations to identify and select appropriate configuration options that optimize performance under specific resource limits. Additionally, designing and generating synthetic workloads for iterative evaluations is crucial to validate the effectiveness of chosen configurations under varied conditions \cite{lee2024k2vtune}. To avoid performance degradation due to evolving workload demands, experts must consider and continuously adapt and modify configuration settings. Achieving this comprehensive tuning workflow effectively and efficiently necessitates sophisticated tools and methodologies that streamline each step, minimizing reliance on manual intervention and expert intuition \cite{jia2020kill, cao2020characterizing}.

Automated tuning solutions for LSM-KVS utilize methods such as reinforcement learning \cite{mo2023learning}, active learning \cite{yu2024camal}, and Bayesian optimization \cite{10.1145/3437984.3458841}, to suggest configurations, minimizing the need for manual adjustments. While effective in specific scenarios, these methods often focus on narrow subsets of parameters and struggle with the interdependency within LSM-KVS components. For instance, frameworks like Endure \cite{Endure} target read optimization under static workloads, while RTune \cite{RTune} prioritizes compaction strategies. Furthermore, their reliance on structured data limits their ability to integrate unstructured knowledge from documentation or expert insights. Recently, a new method to break through the subset paradigm has shown promise. ELMo-Tune \cite{Thakkar2024can} is an early endeavor incorporating LLM into LSM-KVS tuning through an iterative feedback loop that integrates prompt construction and system monitoring. While novel, the design fails to address an end-to-end solution that can automate LSM-KVS characterization and modeling. Further, the ELMo-Tune design assumes static workload patterns (e.g., "100\% random read workload"), an improbable scenario in real-world workloads that show shifts in characteristics like key access and query distribution patterns \cite{cao2020characterizing}.

An ideal solution in this domain demands three essential qualities: 1) a comprehensive understanding of hardware, system, resource, and workload characteristics that can be advantageous for tuning; 2) the ability to craft LSM-KVS configurations that extend beyond a narrow set of options by leveraging prior knowledge; and 3) the flexibility to efficiently manage resources and adapt to runtime changes in response to dynamically shifting workloads. Modern LLM demonstrates human-like contextualization ability by co-relating and drawing inferences from vast training data across diverse domains, demonstrating the potential to fulfill these qualities. 

However, integrating LLM into an end-to-end LSM-KVS optimization solution necessitates overcoming four key challenges: 1) How can two fundamentally different technologies—LLMs, which use natural language, and LSM-KVS systems, which rely on strict code-based languages—be effectively integrated? 2) How can LLMs be leveraged to accurately characterize workloads and enable self-configurable benchmarking? 3) What approaches can be used to design a robust tuning framework capable of managing the complex interdependency among numerous parameters? and, 4) How can real-time system adaptation to unpredictable workload changes be achieved using LLM-driven insights?

To address these challenges, we propose \textbf{ELMo-Tune-V2}, a novel LLM-centric end-to-end optimizing framework. ELMo-Tune-V2 leverages the contextual and generative capabilities of LLM to automate workload characterization, modeling, and tuning for LSM-KVS systems. ELMo-Tune-V2 incorporates three core innovations: 1) synthetic workload generation via LLM-based trace analysis and characterization, 2) an iterative tuning framework with fine-tuning to facilitate macro and micro changes to the LSM-KVS configurations, and 3) real-time dynamic tuning capabilities to respond to evolving workload patterns. 

ELMo-Tune-V2 generates highly adaptable LSM-KVS benchmarks from upper-layer applications by extracting critical workload characteristics from LSM-KVS traces, such as query distribution and access patterns, and transforming them into a flexible JSON-based modeled benchmark. It combines statistical analysis with LLM-driven pattern recognition to capture diverse workload behaviors, extending beyond traditional methods that necessitate manual curve selection and fitting. Next, tuning is performed in ELMo-Tune-V2, similar to how a human expert iteratively performs tuning - modifying a small set of options and then fine-tuning them for greater effectiveness. The modeled benchmark is iteratively tuned to maximize LSM-KVS performance metrics (e.g., increasing throughput, decreasing p99 latency) with ELMo-Tune-V2 generating prompts that contain LSM-KVS configuration, LSM-KVS telemetry (e.g., query distribution), and system resource (e.g., CPU, memory) utilization statistics. The tuning approach is responsible for making changes to the entire LSM-KVS configuration space, and the fine-tuning mechanism forms an inner loop in ELMo-Tune-V2 to refine selected configurations incrementally. ELMo-Tune-V2 implements real-time tuning to maintain performance for unanticipated workload shifts. By continuously monitoring resource utilization and workload characteristics, ELMo-Tune-V2 prompts the LLM about the pattern shift and updates runtime mutable LSM-KVS configurations, such as background job limits or query priorities, without downtime. 

We implement a prototype of ELMo-Tune-V2 using Python and RocksDB and open-source our code on GitHub\footnote{https://github.com/asu-idi/ELMo-Tune-V2}. We test ELMo-Tune-V2's effectiveness across diverse scenarios, including real-world applications (NebulaGraph \cite{wu2022nebula}, Kvrocks \cite{apache_kvrocks}), macro (MixGraph \cite{cao2020characterizing}, YCSB \cite{ycsb}), and micro-benchmarks (varying read-write ratios), finding an overall end-to-end performance uplift of 26-34\% over default settings and 446\%-1485\% over macro and micro-benchmarks. Sensitivity analyses reveal improvements of up to 238\% across different storage devices and workloads. Further, we perform similarity tests for ELMo-Tune-V2's characterization and modeling capabilities using statistical modeling with $p$-value, finding high similarity with $p$-value 0.95. Further, we explore the impact of different LLM models, LLM of different parameter sizes, and different LSM-KVS in the tuning process. We also identify scenarios of how different prompting methods in ELMo-Tune-V2 perform. Our results highlight ELMo-Tune-V2's adaptability and provide actionable insights into tuning strategies for other data systems.

\section{Background}
\label{sec:background}

\begin{figure}[t]
    \centering
    \includegraphics[width=\linewidth]{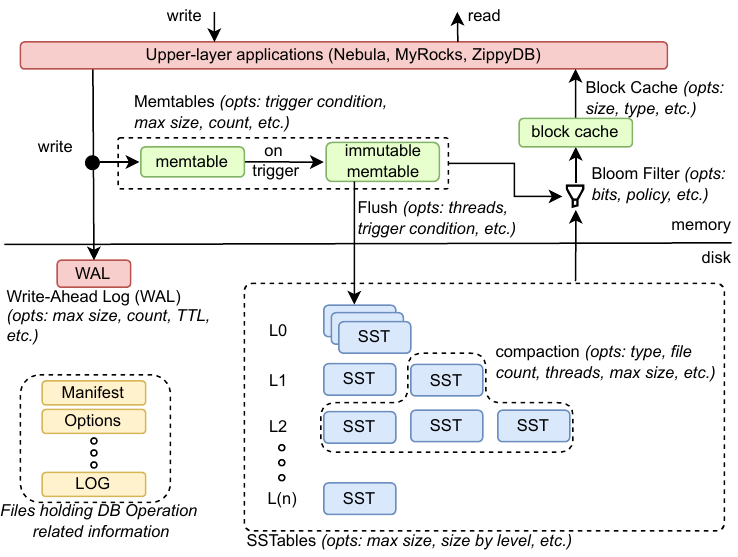}
    \caption{Different Components and their Configurable Options in LSM-KVS.} 
    \label{fig:lsm-kvs-design}
\end{figure}

\subsection{LSM-KVS}
Log-Structured Merge-tree-based Key-Value Stores (LSM-KVS), such as RocksDB \cite{rocksdb}, LevelDB \cite{leveldb}, Cassandra \cite{cassandra}, and HBase \cite{hbase}, form a foundational component to store different types of data (e.g., relational data \cite{matsunobu2020myrocks} or semi-structured data \cite{chang2008bigtable}) and support various modern applications (e.g., big data analytics \cite{wang2013plsm}, artificial intelligence \cite{cao2020characterizing}, and search services \cite{bing_rocksdb_2021}). LSM-KVS employs an append-only, out-of-place update strategy to persistent key-value pairs in files called Sorted String Tables (\textit{SSTables}) combined with in-memory buffering in self-sorting structures (e.g., skiplists) called \textit{memtables} to facilitate large-scale data ingestion with a high-throughput. Additionally, LSM-KVS possess modular components (shown in Figure \ref{fig:lsm-kvs-design}) including memtables, \textit{Write-Ahead-Log} (WAL), \textit{compaction} (a process to perform SST file merge sort across levels), \textit{flush} (persist memtables), and \textit{block cache} \cite{rocksdb}.

Popular LSM-KVS like RocksDB provide over 100 tunable parameters for configuring and tuning the LSM-KVS to different use cases \cite{rocksdb}. Appropriate tuning of these configurations heavily depends on the hardware deployments (e.g., different types and sizes of CPU, memory, and storage \cite{yu2024caas-lsm}) and workloads (e.g., from transactional to analytic queries \cite{cao2020characterizing}). Moreover, tuning is complicated by the interdependence of parameters \cite{dong2021evolution}, as they often compete for the same pool of CPU, memory, and storage resources (e.g., foreground queries competing with background operations like Compaction and Flush), making it challenging to extract the best performance from LSM-KVS. The end-to-end optimization of such LSM-KVS is a multi-stage process \cite{cao2020characterizing}: 1) starting with workload characterization - understanding the underlying hardware and the high-level application workload interaction with the LSM-KVS, 2) followed by modeling the workload (even further developing the benchmarks) for offline testing, and 3) the modeled benchmark is utilized for iterative tuning of configurations before the configurations are deployed in production (e.g., AB-tests or shadow tests) to evaluate and adjust the final tuned configuration \cite{cao2020characterizing}.

\subsection{Hardware, System, and Resource Dependency of LSM-KVS}

\begin{figure}[t]
    \centering
    \includegraphics[width=\linewidth]{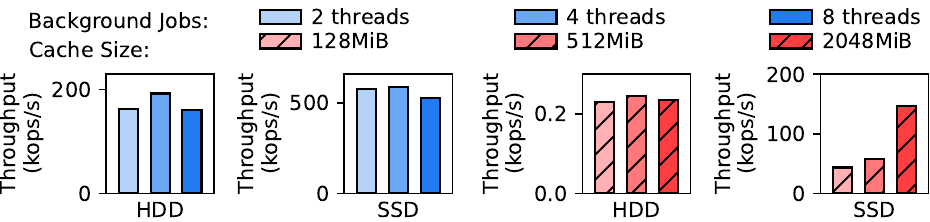}
    \begin{subfigure}{0.47\linewidth}
        \centering
        \caption{Random Write with Varying Background Jobs}
        \label{fig:lsm-kvs-perf-fillrandom}
    \end{subfigure}
    \hfill
    \begin{subfigure}{0.47\linewidth}
        \centering
        \caption{Random Read with Varying Cache Sizes}
        \label{fig:lsm-kvs-perf-readrandom}
    \end{subfigure}
    \caption{LSM-KVS performance on different Hardware, Workload, and Configuration.} 
    \label{fig:lsm-kvs-perf}
\end{figure}

With LSM-KVS embedded in modern applications (e.g., MyRocks \cite{matsunobu2020myrocks}, Kvrocks \cite{apache_kvrocks}, or Nebula Graph \cite{wu2022nebula}), optimizing its performance is crucial. LSM-KVS optimization involves untangling complex dependencies between hardware capabilities, workload patterns, and resource availability.
The underlying storage \textit{\underline{hardware}} plays a significant role in shaping performance. For example, Figure \ref{fig:lsm-kvs-perf-fillrandom} shows the random write throughput on a high-bandwidth SAS SSD significantly outpaces that of a SATA HDD, which is hindered by frequent write stalls. Similarly, Figure \ref{fig:lsm-kvs-perf-readrandom} demonstrates SSD delivers superior random read throughput due to its higher bandwidth and lower latency against HDD. This indicates that HDD workloads require additional resources to mitigate I/O bottlenecks, such as increasing the number of background threads for compaction and flush operations and increasing the block cache size to reduce the storage system reads.

However, increasing \textit{\underline{resources}} haphazardly can lead to over-provisioning and degraded performance. Figure \ref{fig:lsm-kvs-perf-fillrandom} shows that while increasing background threads initially boosts throughput for HDDs and SSDs, excessive background jobs lead to lock contention between threads, resulting in performance drops. This underscores the intricacies of tuning, where indiscriminate scaling of resources may negate its intended benefits. Tuning introduces additional challenges when considering the different behaviors of different configuration parameters. For example, as shown in Figure \ref{fig:lsm-kvs-perf-readrandom}, increasing cache size consistently improves random read performance for both HDDs and SSDs by reducing I/O dependency, demonstrating that certain resources with certain configurations have a positive and predictable impact on performance. 

\begin{figure*}
    \centering
    \includegraphics[width=\linewidth]{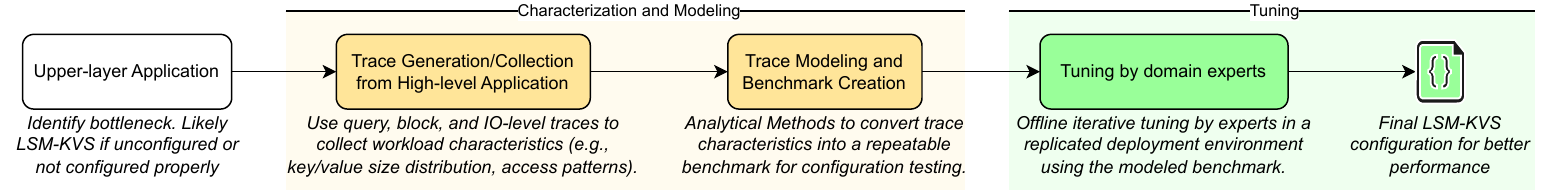}
    \caption{Workflow to achieve performant tuning configuration for LSM-KVS}
    \label{fig:lsm-kvs-tuning}
\end{figure*}

\subsection{Characterizing \& Modeling Workloads for LSM-KVS} \label{sec:modeling}

Domain experts follow a multi-stage process to achieve the end-to-end optimized and fine-tuned configuring for LSM-KVS \cite{cao2020characterizing}, as shown in Figure \ref{fig:lsm-kvs-tuning}. They start by monitoring the overall system to identify bottlenecks that may arise from default or unoptimized configurations of LSM-KVS. The workload is traced and analyzed with detailed characteristics like key-access distribution, query distributions, and key/value size distribution \cite{Endure}. Following this, the workload characteristics are modeled using statistical methods to recreate a benchmark for synthetic workload generation that can be iteratively tested offline when the domain experts conduct the configuration tuning \cite{cao2020characterizing}. Once the experts achieve the desired LSM-KVS performance, the updated configurations are implemented in the production for final calibration. 

Workload characterization is the first stage for tuning LSM-KVS. This process is carried out through performance profiling and trace collection, capturing the metrics on different levels from the applications on LSM-KVS. For example, in RocksDB, a widely used LSM-KVS, three levels of trace collecting mechanisms are implemented with varying overheads \cite{rocksdb}, including Query-Level Traces, Block-Cache Level Traces, and I/O-Level Traces. Those traces are analyzed by experts with profiling tools, statistical analyses, and visualization techniques to provide insights, such as identifying workload hotspots, imbalances in read/write ratios, and storage media bottlenecks. In particular, Query-Level Traces (QLTs) are widely used to identify application key-value workload characteristics, including access patterns, skewness, and popularity distributions. These insights create modeled workloads for benchmarking and tuning the LSM-KVS \cite{lee2024k2vtune, cao2020characterizing}.

Modeling then focuses on closely emulating workload characteristics within benchmarks, enabling iterative tuning on alternative machines for offline testing of the LSM-KVS. Recent advances in machine learning (ML) have facilitated automated workload characterization and modeling in domains like networking \cite{wang2017machine} and high-performance computing \cite{tuncer2017diagnosing}, leveraging synthetic trace generation and hybrid ML techniques to minimize the overhead of manual tracing \cite{paul2022machine}. However, the LSM-KVS domain remains unexplored mainly due to the complexity of LSM-KVS workload patterns that feature very dynamic access patterns, a wide array of possible deployments, and key-value-related characteristics.

\subsection{Configuring and Tuning LSM-KVS}

Widely used LSM-KVS systems such as RocksDB and HBase offer over 100 configurable options to tailor key functions like compaction strategies, memory allocation, and I/O scheduling \cite{rocksdb,hbase}. 
The interplay between these parameters and their significant quantity introduces significant complexity, demanding deep domain expertise to achieve optimal configurations. This complexity is compounded by the dynamic nature of workloads, where suboptimal tuning can lead to severe performance degradation, further highlighting the challenges in balancing resource utilization effectively.

Historically, this process is performed manually in an iterative manner by domain experts who validate changes using modeled benchmarks to achieve optimal performance tuning for the expected application workloads, as described in Section \ref{sec:modeling}. However, this process is resource-intensive, time-consuming, and error-prone, especially in an ever-expanding space of tunable options in LSM-KVS \cite{dong2021evolution}. Automated alternatives leverage Machine Learning (ML) and optimization algorithms to address this problem. For instance, RTune \cite{RTune} integrates deep learning and genetic algorithms to forecast optimal configurations by analyzing workload patterns, Dremel \cite{Dremel} adopts a Multi-Armed Bandit model for online configuration selection, CAMAL \cite{yu2024camal} leverages active learning to perform decoupled tuning of parameters like size ratios between adjacent levels, memory allocations to write buffer and bloom filters, and ADOC \cite{yu_adoc_2023} modifies the LSM-KVS source code to automatically adjusts block sizes and background resource availability to harmonize data flow among LSM-KVS components. However, these automated approaches cater to a limited subset of configuration options and lack a comprehensive understanding of system resources and workloads. Moreover, these methodologies cannot leverage information from unstructured insights, such as online documentation, blogs, websites, or expert-provided heuristics.

Recently, a new domain that bridges the gap to leveraging unstructured insights has emerged as Large Language Model-based (LLM-based) tuning frameworks. Unlike traditional analytical methods or ML models that rely on structured data, LLMs can derive insights from diverse sources, such as source code, logs, configuration files, or expert-written best practices. This flexibility positions LLMs as a promising approach to addressing the complexity and interdependency inherent in LSM-KVS workloads and configurations. Early efforts to apply LLMs to data system tuning, such as the ELMo-Tune framework \cite{Thakkar2024can} for LSM-KVS and $\lambda$-tune \cite{giannankouris2024lambda, giannakouris2024demonstrating} for OLAP applications (e.g., Postgres), have demonstrated the potential of more effective tuning. In particular, ELMo-Tune attempts to emulate a human, employing a feedback loop comprising modules for prompt generation, LSM-KVS configuration, and safeguard enforcement to adjust configurations based on system performance and predefined workloads iteratively.

\section{Motivation and Challenges}
\label{sec:motiv-challenges}

\begin{figure*}
    \centering
    \includegraphics[width=\linewidth]{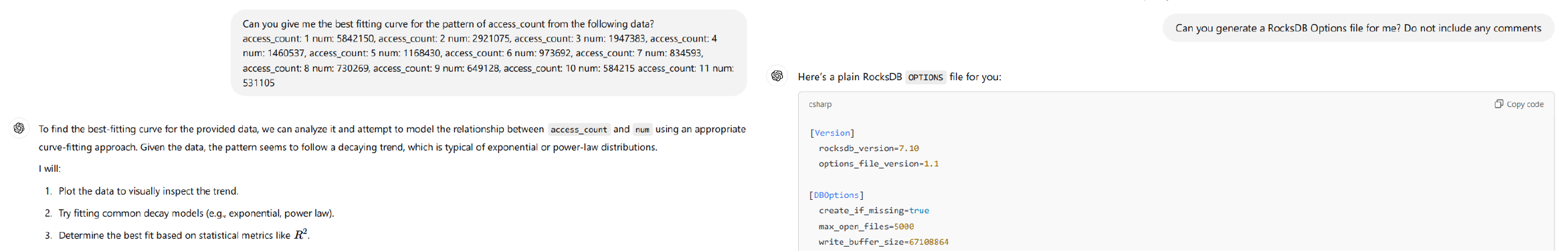}
    \begin{subfigure}{0.47\linewidth}
        \centering
        \caption{Curve Fitting of Trace Analysis from LSM-KVS Trace}
        \label{fig:motiv-analysis-gpt}
    \end{subfigure}
    \hfill
    \begin{subfigure}{0.47\linewidth}
        \centering
        \caption{Generation of LSM-KVS configuration file}
        \label{fig:motiv-generation-gpt}
    \end{subfigure}
    \caption{LLM Generational and Analytical Capability.}
    \label{fig:motiv-gen-analysis-gpt}
\end{figure*}

The end-to-end optimization of LSM-KVS as a whole remains largely unexplored. Individually, workload characterization and modeling solutions for LSM-KVS are sparse, and tuning solutions limit themselves to a subset of options \cite{yu_adoc_2023, RTune, yu2024camal, lee2024k2vtune, Endure} or assume a fixed workload pattern \cite{Thakkar2024can}.
An ideal solution in the domain necessitates three important characteristics: 
1) Understanding of workload characteristics, in conjunction with hardware, system, resource knowledge, 
2) Ability to construct LSM-KVS configuration beyond a subset of options by leveraging the pre-existing knowledge, and 
3) Adaptivity to effectively consume resources and perform runtime changes for dynamically changing workloads. 

\subsection{Motivation}

Modern LLMs, such as GPT-like architectures, are designed based on the transformer architecture \cite{vaswani2017attention} and are trained with high-quality datasets containing structured data (e.g., JSON, telemetry, metrics) and unstructured data (e.g., wikis, code repositories, forums, manuals) \cite{meta2025llama3}. 
It enables models to excel at tasks such as reasoning, summarization, and creative text generation \cite{malach2023auto}, which aligns with LSM-KVS optimization needs. First, the large training corpus enables LLM to synthesize patterns and insights (shown in Figure \ref{fig:motiv-analysis-gpt}), modeling complex workloads like those in LSM-KVS systems using hardware, system, and resource data. Second, their natural language processing capabilities move beyond the subset configuration paradigm of existing solutions for LSM-KVS, providing the full configuration space exploring capability, shown in Figure \ref{fig:motiv-generation-gpt}. Finally, by analyzing temporal LSM-KVS telemetry, LLMs can act as real-time agents, proactively adapting to performance drops or workload shifts.

\vspace{0.3em}
\noindent\textbf{Leveraging LLM Cross-Domain Knowledge for Automated Workload Characterization and Modeling.} 
Traditionally, workload characterization and modeling in an LSM-KVS system relies on manual expert interpretation of collected traces (e.g., QLT). This process is time-consuming and requires experts to analyze patterns (e.g., Zipfian, Gaussian) and match them to the closest fitting pattern. Such an approach heavily depends on domain expertise and is not easily scalable. Existing automated trace analyzing \cite{cao2020characterizing} are sparse and limited to predefined patterns, which fail to capture periodic or unprogrammed workload behaviors. LLMs provide a flexible and scalable solution \cite{naveed2023comprehensive}. By synthesizing insights from their extensive training data, which spans a wide range of behavioral patterns, and utilizing the attention mechanism to detect periodicity, they can emulate human experts in identifying nuanced patterns \cite{rosin2022temporal}. This capability allows LLMs to autonomously perform workload characterization without requiring predefined workload specifications. As a result, LLMs create a more resilient, automated system capable of adapting to evolving workload patterns, ultimately improving efficiency and accuracy. 

\vspace{0.3em}
\noindent\textbf{Tuning LSM-KVS Beyond the Subset Paradigm.}
Traditional tuning approaches for LSM-KVS systems often limit optimization to a subset of configuration parameters. These methods rely heavily on machine learning (ML) \cite{RTune, Dremel, 10.1145/3437984.3458841} or heuristic-based \cite{rocksdb_tuning_guide, intel_rocksdb_xeon_tuning} approaches and are constrained by their inability to consider the broader context of hardware, resource, and software. As a result, they leave significant potential for performance gains untapped. In contrast, LLMs, trained on extensive and diverse datasets, can capture intricate relationships from diverse domains. This can be leveraged in tuning LSM-KVS, where hardware characteristics, workload patterns, and resource availability all play a role and must be understood for best tuning. Further, LLMs are not bound to any subset of options; the vast training dataset enables a comprehensive exploration of the configuration space, accounting for the co-dependencies and trade-offs between foreground and background operations. This approach allows for more context-aware optimization that goes beyond the limitations of traditional methods, providing a path to more effective and scalable performance tuning.

\vspace{0.3em}
\noindent\textbf{LLM-Driven Real-Time Tuning for Adaptability.}
The dynamic nature of workloads introduces additional complexity to the tuning process, as systems must continuously adjust their configurations to meet shifting demands. Current automated tuning approaches are often limited by their focus on specific parameters (e.g., ADOC \cite{yu_adoc_2023}) or their reliance on workloads with similar characteristics (e.g., Endure \cite{Endure}). These methods are insufficient for real-time tuning, particularly when handling more workloads that show unanticipated shifts in patterns. LLMs provide a solution to this with their ability to decipher patterns. When detecting a shifting pattern (e.g., key/value sizes, query distribution) for analysis of LSM-KVS telemetry, the LLM can perform prompt execution of real-time modifications to the runtime mutable options of LSM-KVS. The analysis and generative capabilities of LLM can be leveraged in real-time tuning to not only enhance performance but also minimize the operational overhead associated with manual interventions.

\subsection{Research Objectives and Challenges}
\label{sec:research-obj-challenges}

Motivated by the potential of LLMs, the primary objective of this research is to design an LLM-driven framework that delivers comprehensive optimization for LSM-KVS systems, spanning the entire life cycle of workload characterization, modeling, and system tuning. 
Four primary challenges must be resolved to achieve this objective. 
\textit{\underline{First}}, creating a \textbf{framework that integrates LLM and LSM-KVS}. How to create a framework that leverages the cross-domain capabilities and derive insights from the vast training data that the LLM is trained on while also being able to interact with the LSM-KVS seamlessly? An error or hallucination from the LLM should not hamper the system and should be resolved automatically.
\textit{\underline{Second}}, develop a \textbf{dynamic workload characterization process} that models the workload and can capture real-world temporal and spatial variability. How to leverage LLMs to achieve automatic benchmark creation based on the workload characteristics? 
\textit{\underline{Third}}, design a \textbf{comprehensive tuning framework} that supports both coarse-grained configuration and fine-grained adjustments. This framework must account for intricate interdependency between parameters while enabling minute changes to those parameters to extract the best performance. How to involve LLM assistance to navigate the vast configuration space and deliver optimizations at both macro and micro levels?
\textit{\underline{Fourth}}, enable \textbf{real-time tuning of LSM-KVS} systems that can adapt to unanticipated workload changes in real time. How can the domain knowledge of LLM be utilized for the more niche problem of tuning dynamically modifiable LSM-KVS options if a workload breaks pattern or resources are underutilized?

\begin{figure*}
    \centering
    \includegraphics[width=\linewidth]{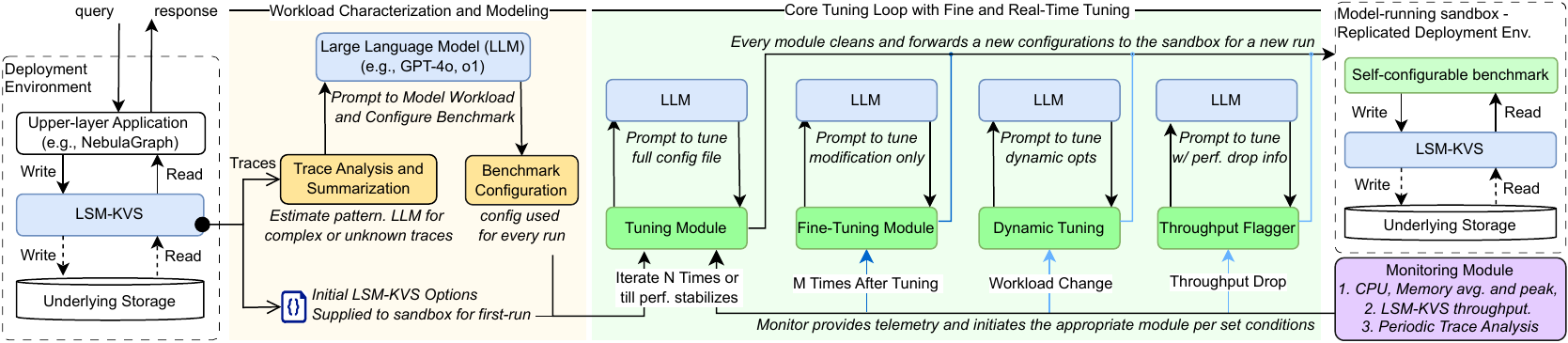}
    \caption{ELMo-Tune-V2 Framework}
    \label{fig:ELMo-Tune-V2-framework}
\end{figure*}

\section{ELMo-Tune-V2 Design}
\label{sec:design}

To solve the challenges from Section \ref{sec:research-obj-challenges}, we present ELMo-Tune-V2, which harnesses the potential of an LLM to perform automatic workload characterization, modeling, and iterative tuning of LSM-KVS with different hardware and application workloads. It leverages Query-Level Trace (QLT) and integrates knowledge of hardware, software, and workload characteristics to go beyond traditional subset-based optimization. ELMo-Tune-V2 offers offline and real-time tuning capabilities and can perform fine-grained adjustments to optimize system performance.

\subsection{Architecture Overview and Methodology}

ELMo-Tune-V2, illustrated in Figure \ref{fig:ELMo-Tune-V2-framework}, comprises three interconnected modules: 
1) The Workload Characterization and Self-Configured Benchmark Module processes QLTs to extract upper-layer workload signatures using an approach that combines LLM-guided analysis with deterministic curve-fitting. On pattern identification, ELMo-Tune-V2 prompts the LLM to transform the workload characteristics into a benchmark configuration file used by the tuning module. 
2) The Tuning Engine, in conjunction with the Fine-Tuning module, iteratively utilizes LLM to suggest LSM-KVS parameter adjustments both on a file and subset level aimed at improving system performance metrics (e.g., throughput and p99) by making macro and micro changes. The tuning loop incorporates the LLM for LSM-KVS parameter adjustments to improve system performance metrics. The internal fine-tuning loop enables the LLM to focus on making smaller changes to a subset of options selected for modification by the tuning module.
And 3) The Real-Time Tuning Module extends the Tuning Engine's functionality by dynamically adjusting modifiable LSM-KVS parameters during runtime. This ensures adaptability to changing workloads and continuous system performance optimization in real-time.

\subsection{LLM-Assisted Workload Characterization and Self-Configured Benchmark}

Insights from QLT, such as key-value access patterns, query distributions, and key/value size distributions, can be refined into actionable patterns that facilitate workload replication and benchmark creation for system tuning. However, this process is challenging due to the reliance on user expertise in existing curve-fitting tools like SciPy \cite{virtanen2020scipy} and MATLAB \cite{betzler2003fitting}, which necessitate the user to define fitting patterns. A promising solution lies in leveraging an LLM, which can synthesize insights from its extensive training data and correlate them with observed QLT data. This approach enables more flexible and adaptive curve fitting, unlocking insights into workload characteristics that traditional methods might overlook.

\begin{figure}
    \centering
    \includegraphics[width=0.95\linewidth]{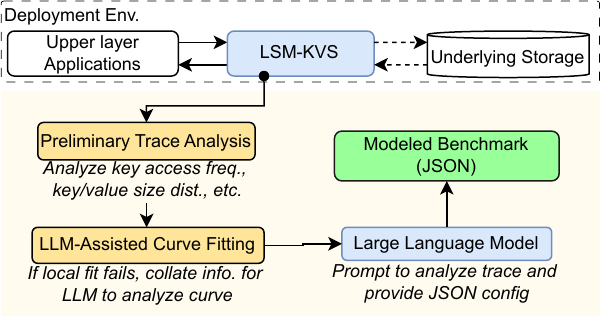}
    \caption{Characterization, Modeling and Benchmark Configuration in ELMo-Tune-V2}
    \label{fig:ELMo-Tune-V2-characterization-framework}
\end{figure}

However, directly feeding millions of QLT records into an LLM exceeds its context length constraints, necessitating processing. ELMo-Tune-V2 introduces a novel solution to address this, as illustrated in Figure \ref{fig:ELMo-Tune-V2-characterization-framework}. ELMo-Tune-V2 combines pre-prompting analysis to identify initial patterns and refines them through LLMs, offering better fits when preliminary analysis alone falls short. Furthermore, ELMo-Tune-V2 automates benchmark synthesis that emulates the identified workload characteristics, enabling repeatable and iterative tuning of LSM-KVS. ELMo-Tune-V2 features a JSON-based interface to provide a common, flexible interface for benchmark synthesis, ensuring seamless communication and efficient configuration generation.

ELMo-Tune-V2 first analyzes the trace, extracting workload characteristics such as the mean, median, and mode of key and value sizes, along with the patterns of key size, value size, and access distribution. To determine the best-fitting curve for key-size, value-size, and access pattern distributions, a pre-determined set of common distributions (e.g., Zipfian from Yahoo's YCSB \cite{ycsb}, sum of two exponential from Meta's MixGraph \cite{cao2020characterizing}) are evaluated using statistical methods like the coefficient of determination ($R^2$) to find the best-fitting curve. However, ELMo-Tune-V2 leverages its LLM integration to consider a wider set of patterns when the best-fitting curve fails to meet preset statistical thresholds. To stay within the bounds of the LLM context length, ELMo-Tune-V2 transforms the QLT into a time-windowed trace that condenses information into mathematical properties (mean, median, mode key/value size, access counts within the period) for user-set periods. This time-windowed trace file is collated with earlier statistical results information and sent as a prompt to the LLM. The LLM behaves as an expert, leveraging its training on diverse datasets to identify potential candidate curves and suggesting periodic patterns based on the input trace. 

Post characterization, the workload is modeled into a benchmark for iterative tuning. Given the wide variety of possible patterns (e.g., Zipfian, Gaussian, Exponential, etc.) of workload characteristics like key access, key size, and value size distribution, it is pertinent to design a highly configurable solution. Further, LLM integration in characterization necessitates the solution to communicate automatically between the LSM-KVS and the LLM. ELMo-Tune-V2 designs a JSON interface to run multiple workloads one after another and contains information like start time, workload type (e.g., random writes, reads), duration, key/value size distribution (e.g., Zipfian, Pareto), standard deviation in value size, query ratios, and access distributions. To emulate complex real-world workloads that can vary in throughput intensity, we design the benchmark to spawn child threads that utilize the same DB pointer and can emulate multiple clients for workloads that necessitate stress tests with multiple clients/writer threads. Such an approach offers a simple-to-use and highly configurable parameter space while being easily interpretable by an LLM. 

\subsection{Core Iterative Tuning Framework}

\begin{figure}
    \centering
    \includegraphics[width=0.95\linewidth]{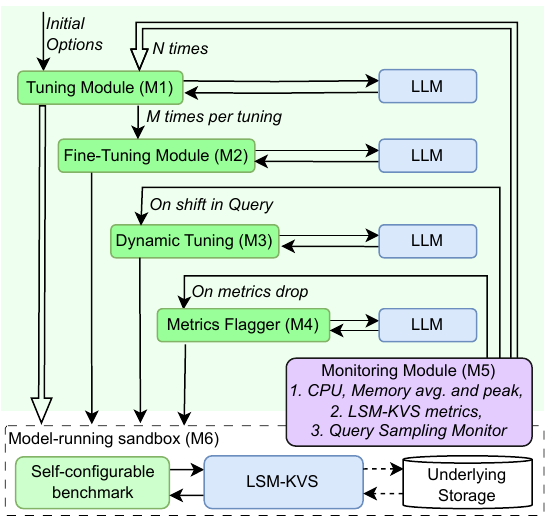}
    \caption{Iterative-Tuning in ELMo-Tune-V2}
    \label{fig:ELMo-Tune-V2-tuning-framework}
\end{figure}

To harness the cross-domain contextualization capabilities of an LLM and explore the vast configuration space of LSM-KVS, we propose an iterative loop framework (shown in Figure \ref{fig:ELMo-Tune-V2-tuning-framework}) for tuning an LSM-KVS. The iterative loop starts with default configurations from the LSM-KVS (e.g., RocksDB default Option file), which are tested in a sandbox environment (see M6 in Figure \ref{fig:ELMo-Tune-V2-tuning-framework}) that is emulated with the same resource (e.g., CPU, Memory) constraints as the targeted production deployments. An LSM-KVS telemetry and resource utilization monitoring module (see M5) collects constant feedback from the sandbox that is fed to the tuning module (see M1). Finally, the tuning module creates prompts and adjusts parameters using a fine-tuning loop (see M2) within each iteration.

The fine-tuning loop iteratively refines the subset of configurations tuned by the tuning module, aiming to maximize system resource utilization. Apart from that, two additional sub-modules are active during benchmark execution. The throughput flagger (see M4) detects performance drops, halts iterations, and prompts the LLM to refine configurations. Simultaneously, a real-time tuning module (see M3) dynamically adjusts configurations to accommodate workload shifts and maximize runtime resource utilization. This tuning continues until stable performance or the user-defined iteration limit is achieved. Further details on the sub-modules are discussed below.

\vspace{0.3em}
\noindent\textbf{Model-running sandbox} 
executes the benchmark scenarios defined by ELMo-Tune-V2's characterization and modeling on the LSM-KVS with different configuration files as provided to the LSM-KVS, simulating operations under controlled CPU and memory conditions targeted by the user.

\vspace{0.3em}
\noindent\textbf{Monitoring module} 
tracks peak and sustained resource utilization (CPU and memory) and LSM-KVS metrics by query sampling (query distribution, LSM-KVS internal listeners \cite{rocksdb}), during benchmark execution.

\vspace{0.3em}
\noindent\textbf{Throughput flagger} 
periodically compares current LSM-KVS performance metrics (e.g., throughput) with historical data from the monitoring module. If a performance degradation is detected, it halts the current iteration, creates a special prompt indicating the performance drop, and provides a list of modifications that led to the performance drop. This allows the LLM to revise the options that led to the performance degradation and re-initiate the iteration.

\vspace{0.3em}
\noindent\textbf{Tuning module} orchestrates prompt creation by aggregating CPU, memory, and throughput statistics from the monitoring module, along with current and past tuning options. The module is responsible for prompt creation, communication with the LLM, and dissection of LLM response into an options file for the LSM-KVS. 

\begin{figure}
    \centering
    \includegraphics[width=\linewidth]{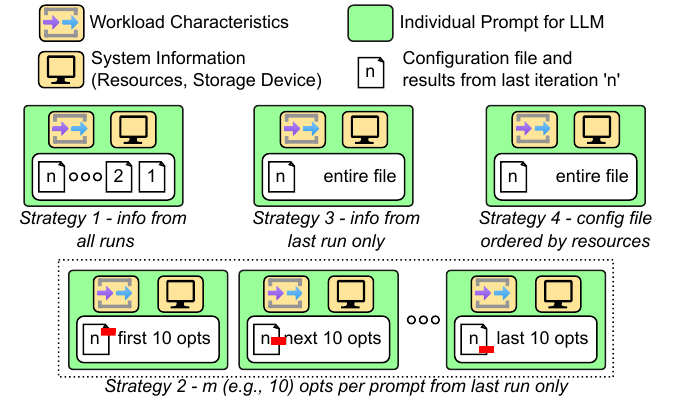}
    \caption{Tuning Prompt Strategies in ELMo-Tune-V2}
    \label{fig:ELMo-Tune-V2-tuning-methods}
\end{figure}

Considering the diverse information available (configuration file, trace details, resource utilization information, previous iteration results) to pass to the LLM, the inclusion and order of content can significantly affect the outcome. To effectively leverage different prompting methodologies, ELMo-Tune-V2 designs four strategies as shown in Figure \ref{fig:ELMo-Tune-V2-tuning-methods} with different purposes - \textit{Strategy 1} forms a straightforward approach that collates and sends all information to the LLM, allowing the LLM to infer freely. A complete list of options and monitoring output from all past iterations is collated and sent as a single prompt. 
The design then includes \textit{Strategy 2} to focus on the LLM to modify more options by dividing the configuration file into multiple parts with only a subset of options each. It collates every subset of options with information from the previous run, prompting the LLM to only make changes within the subset. 
\textit{Strategy 3} takes a simpler human-like approach; the LLM is prompted with information from the latest run's throughput and configuration information. Allowing the LLM a short view into historical information to tune the next run.
Finally, \textit{Strategy 4} focuses on effective resource utilization, dividing the options file into different resource-oriented categories, i.e., options like RocksDB's max\_background\_jobs that utilize majority CPU resources will be classified as such. This method provides the LLM with additional context on the resource reliance of different LSM-KVS configurations, focusing on more effective resource utilization instead of a direct focus on improving LSM-KVS metrics. 

\vspace{0.3em}
\noindent\textbf{Fine Tuning} 
The tuning module is responsible for making large modifications to the LSM-KVS. However, with the many LSM-KVS configuration options, it is important to also focus on and optimize every subset of modified options for best performance or resource efficiency. To achieve this, ELMo-Tune-V2 implements the fine-tuning module that exclusively refines the subset of parameters adjusted by the tuning module. The module targets refining selected parameters with incremental changes to achieve more efficient resource utilization for performance. This is achieved with a prompt that collates from the monitoring module with historical results, prompting the element to generate revised parameters for better resource utilization. This process adds validation against prior throughput from monitoring metrics, ensuring that these refinements result in measurable improvements. 

\vspace{0.3em}
\noindent\textbf{Realtime Tuning}
updates system configurations in real time based on feedback from the monitoring module. Unlike fine-tuning, this sub-module optimizes the system for the current workload, ensuring responsiveness to shifts without downtime. Details are provided in Section \ref{sec:fine-realtime-tuning}.

\subsection{Adaptive Realtime Tuning}
\label{sec:fine-realtime-tuning}

ELMo-Tune-V2 introduces a real-time tuning framework to optimize resource utilization and adapt to unpredictable workloads, addressing a critical gap in existing systems. By periodically adjusting runtime options in LSM-based key-value stores (LSM-KVS) like RocksDB (e.g., max\_background\_jobs), ELMo-Tune-V2 ensures configurations align with dynamic workload conditions such as shifting access patterns or resource constraints. This process is computationally intensive due to the necessity of periodic workload characterization, and further, any large modifications in configurations can take time and additional resources to propagate. For instance, reducing the level size when full can lead to immediate compactions and possibly writing stalls in the LSM-KVS.

ELMo-Tune-V2 periodically checks the resource utilization and the query characterization obtained from workload sampling \cite{rocksdb}. If an unanticipated pattern shift crosses preset thresholds, ELMo-Tune-V2 initiates real-time action by collecting current system information and the new workload pattern and feeding the information into the LLM. The LLM proposes targeted, incremental configuration changes to minimize disruptions and maintain performance. The ideal frequency of these checks varies based on resource availability and workload volatility; for example, if the trace analysis is purely incremental, checking every few seconds will have light overhead; however, if trace analysis is expensive on the setup, invoking the module should be done less periodically. The ELMo-Tune-V2 design acknowledges this and allows the tuning to be user-configurable while providing a default periodicity of once in ninety seconds.

\subsection{Case Study: End-to-End LSM-KVS Optimization for NebulaGraph}

To illustrate the complete end-to-end process of ELMo-Tune-V2, we provide a detailed example using NebulaGraph. NebulaGraph is first configured to use a specific RocksDB options file and modified to enable trace generation during benchmark execution. The LDBC \cite{angles2020ldbc, erling2015ldbc}  benchmark was run to simulate a complex social network workload, generating various queries, including reads, writes, and graph traversals. NebulaGraph recorded detailed traces, capturing workload characteristics such as query distribution, access patterns, and key/value sizes with minimal overhead using an in-memory file system (e.g., tmpfs \cite{snyder1990tmpfs} for low latency. 

ELMo-Tune-V2 then processed these traces to analyze the workload, identifying patterns such as frequently accessed vertices, query hotspots, and read/write ratios, which provided insights into potential performance bottlenecks and resource usage. Based on this workload characterization, ELMo-Tune-V2 generated an optimized RocksDB options file, adjusting key parameters like cache size and compaction strategies to align with the observed workload. The updated configuration was deployed to NebulaGraph (no tracing), and the process was iteratively repeated for a fixed number of iterations. 

\begin{figure*}
    \centering
    \includegraphics[width=\linewidth]{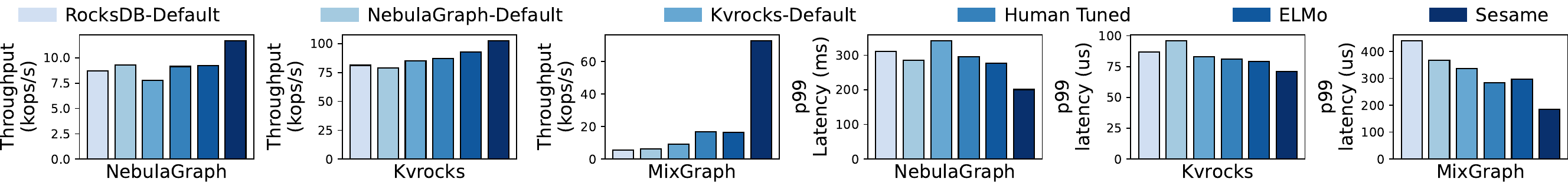}
    \caption{End-to-End Performance of ELMo-Tune-V2 with upper-layer applications (NebulaGraph and Kvrocks) and MixGraph} 
    \label{fig:upper-layer-apps}
\end{figure*}

\begin{figure*}
    \centering
    \includegraphics[width=\linewidth]{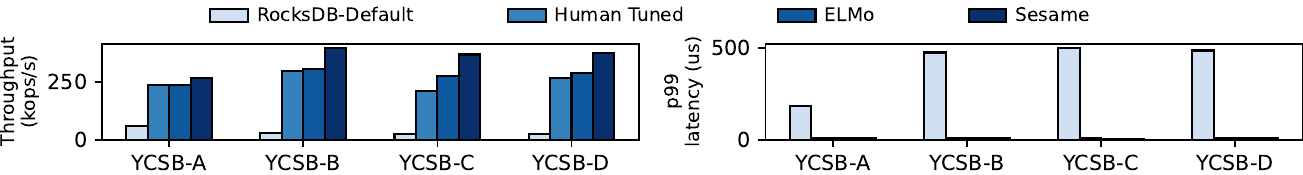}
    \caption{End-to-End Performance of ELMo-Tune-V2 with YCSB Test Suite (A through D)} 
    \label{fig:macro-benchmarks-ycsb}
\end{figure*}

\begin{figure*}
    \centering
    \includegraphics[width=\linewidth]{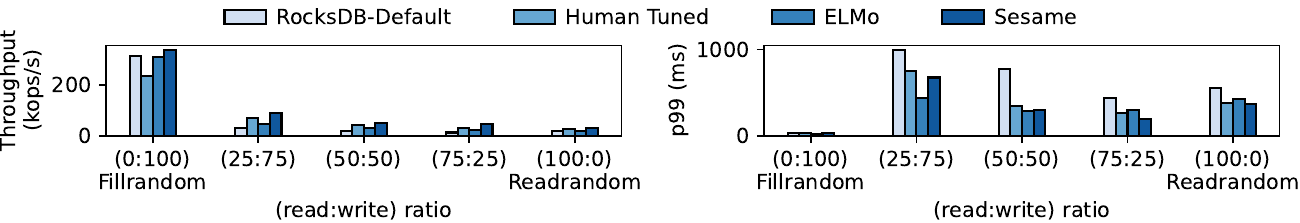}
    \caption{End-to-End Performance of ELMo-Tune-V2 under different read:write ratio workloads} 
    \label{fig:micro-benchmarks}
\end{figure*}

\section{Evaluation and Finding}

We implement ELMo-Tune-V2 using Python (v3.10) \cite{10.5555/1593511} for the core framework, utilizing RocksDB (v8.8.1) \cite{rocksdb_v8_8_1} as the underlying LSM-KVS, and developing the new self-configurable benchmark based on RocksDB's db\_bench \cite{facebook_db_bench}. We utilize OpenAI's o1-preview \cite{openai_gpt-4_2024} API as our LLM agent. The source code for ELMo-Tune-V2 is available on GitHub \cite{open-ELMo-Tune-V2}. 

We conduct a comprehensive evaluation to answer the following questions: 1) Can ELMo-Tune-V2 effectively and consistently outperform current SOTA, human-tuned, and pre-provided default baselines? 2) What is the impact of each module? Is ELMo-Tune-V2 sensitive to changing hardware or workloads? 3) How does ELMo-Tune-V2 utilize prompting? Can the design detect resource under-utilization and implement changes that remedy this? and 4) With the changing landscape of LLMs, how would changing the LLM or using LLM with different parameter sizes affect the system? Further, how would changing LSM-KVS affect the system?  

\subsection{Experimental Setup}

\noindent\textbf{Hardware Setup.}
Our experimental evaluation utilizes three servers:
Server 1 and 2 are identically set up with Intel Xeon Gold 6330 CPU @ 2.00GHz, 256 GiB of RAM, a 3.84TB SAS SSD running Ubuntu 22.04.3 LTS, and Server 3 is set up with Intel Xeon Silver 4310 CPU @ 2.10GHz, 64 GiB of RAM, an 8TB SATA HDD storage running Ubuntu 20.04.6 LTS. We use the identical servers - 1 and 2 - to conduct all our evaluations, except for Section 5.4 and wherever explicitly mentioned. We leverage Linux cgroups \cite{rosen2013resource} on all servers to ensure resources are limited to 4CPU cores and 4GiB of RAM available for any test unless explicitly mentioned.

\vspace{0.3em}
\noindent\textbf{Real-World Applications and Workloads.}
We select two popularly used real-world applications that use RocksDB as their embedded storage engines - NebulaGraph \cite{wu2022nebula} and Kvrocks \cite{apache_kvrocks} to conduct end-to-end evaluation. Kvrocks was evaluated using the Redis benchmark \cite{redis_benchmark}, while NebulaGraph was tested with the LDBC graph benchmark \cite{angles2020ldbc, erling2015ldbc}. For Kvrocks, the Redis benchmark was used to generate SET and GET queries, simulating random read and write operations. The benchmark issued 50 million SET and GET queries using 16 threads. Each key-value pair had a size of 144 bytes, with a key-space range of 10 billion. For NebulaGraph, we utilized the LDBC Social Network Benchmark. First, we generated an SF3-scale graph dataset (10GB), imported it into NebulaGraph, and then executed the Nebula benchmark \cite{nebula_benchmark}.

Further, we use different test cases from popular benchmarking tools, including YCSB (A, B, C, D) and db\_bench consisting of fillrandom, readrandom, readrandomwriterandom, and mixgraph (real-world LSM-KVS workload from Facebook) \cite{cao2020characterizing}. All pure write tests are conducted over 50 Million (M) KV-pairs unless specified, and all tests involve reading over 10M KV-pairs. We utilize eight writer threads to ensure the system is under pressure. Additionally, we enable direct I/O for compaction, reads, and flushes across all tests.

\begin{figure*}
  \begin{minipage}[b]{0.245\textwidth}
    \centering
    \includegraphics[width=\textwidth]{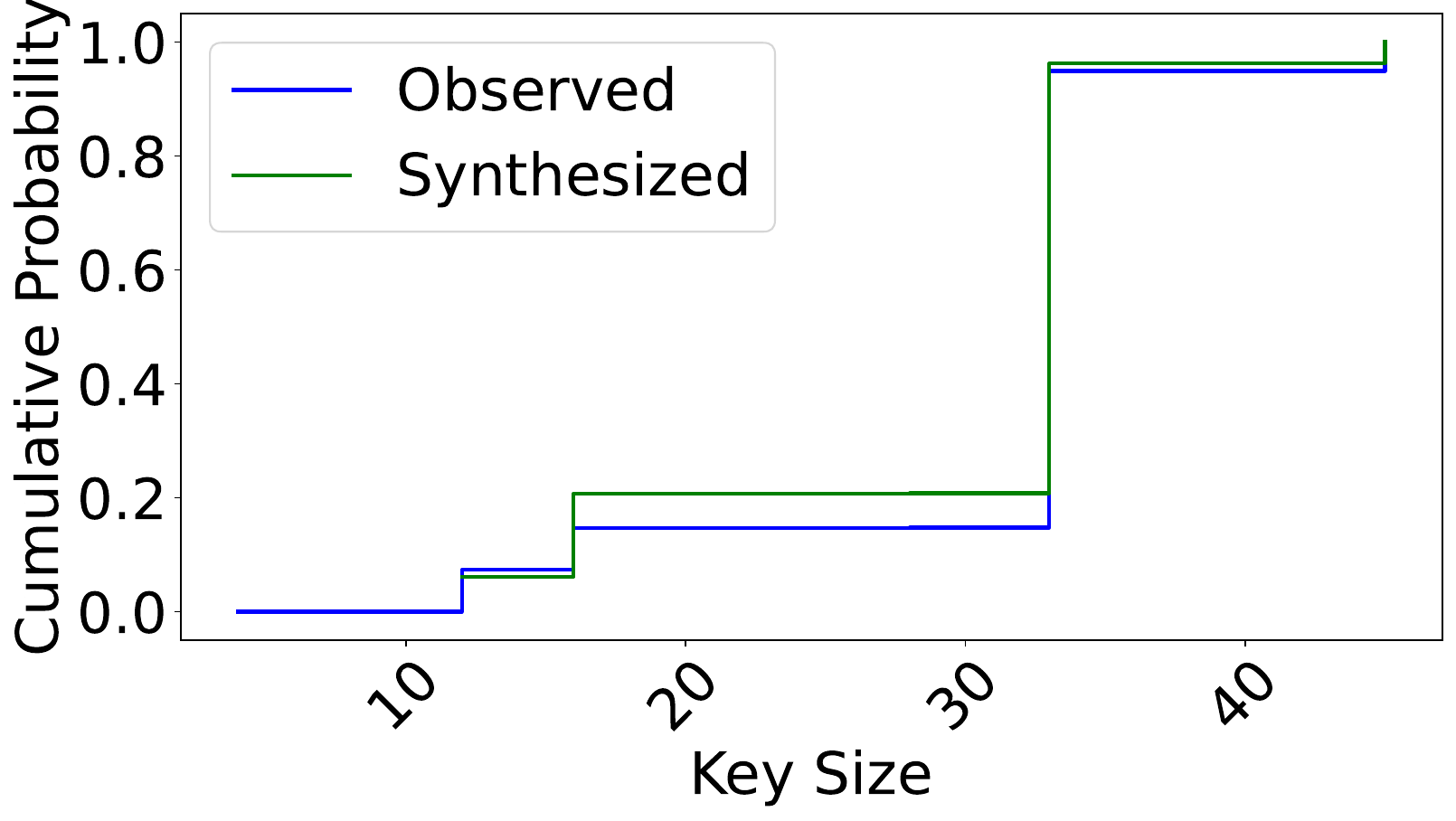}
    \subcaption{Key distribution}
    \label{subfig:key_dis}
  \end{minipage}
  \hfill
  \begin{minipage}[b]{0.245\textwidth}
    \centering
    \includegraphics[width=\textwidth]{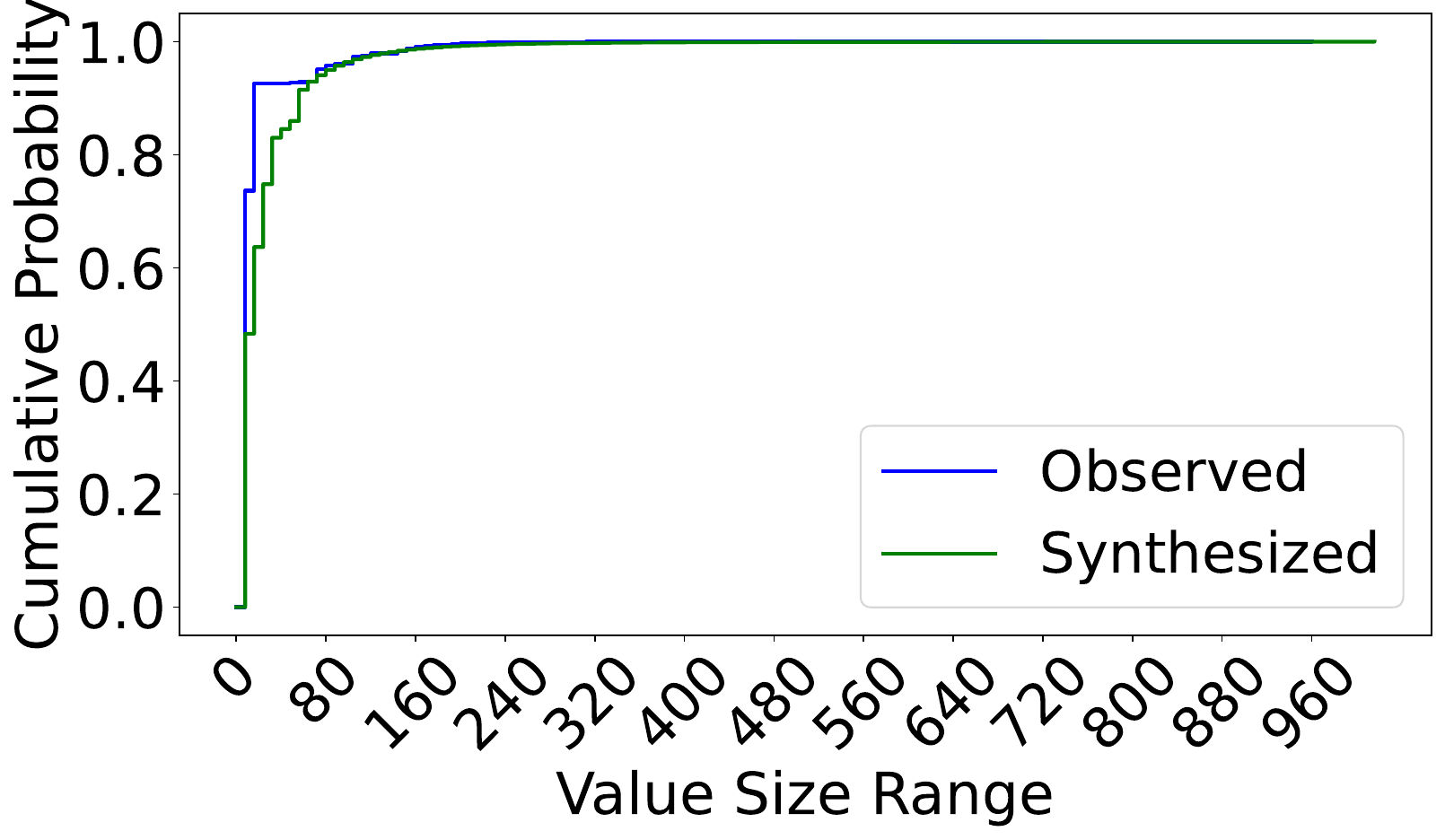}
    \subcaption{Value distribution}
    \label{subfig:value_dis}
  \end{minipage}
  \hfill
  \begin{minipage}[b]{0.245\textwidth}
    \centering
    \includegraphics[width=\textwidth]{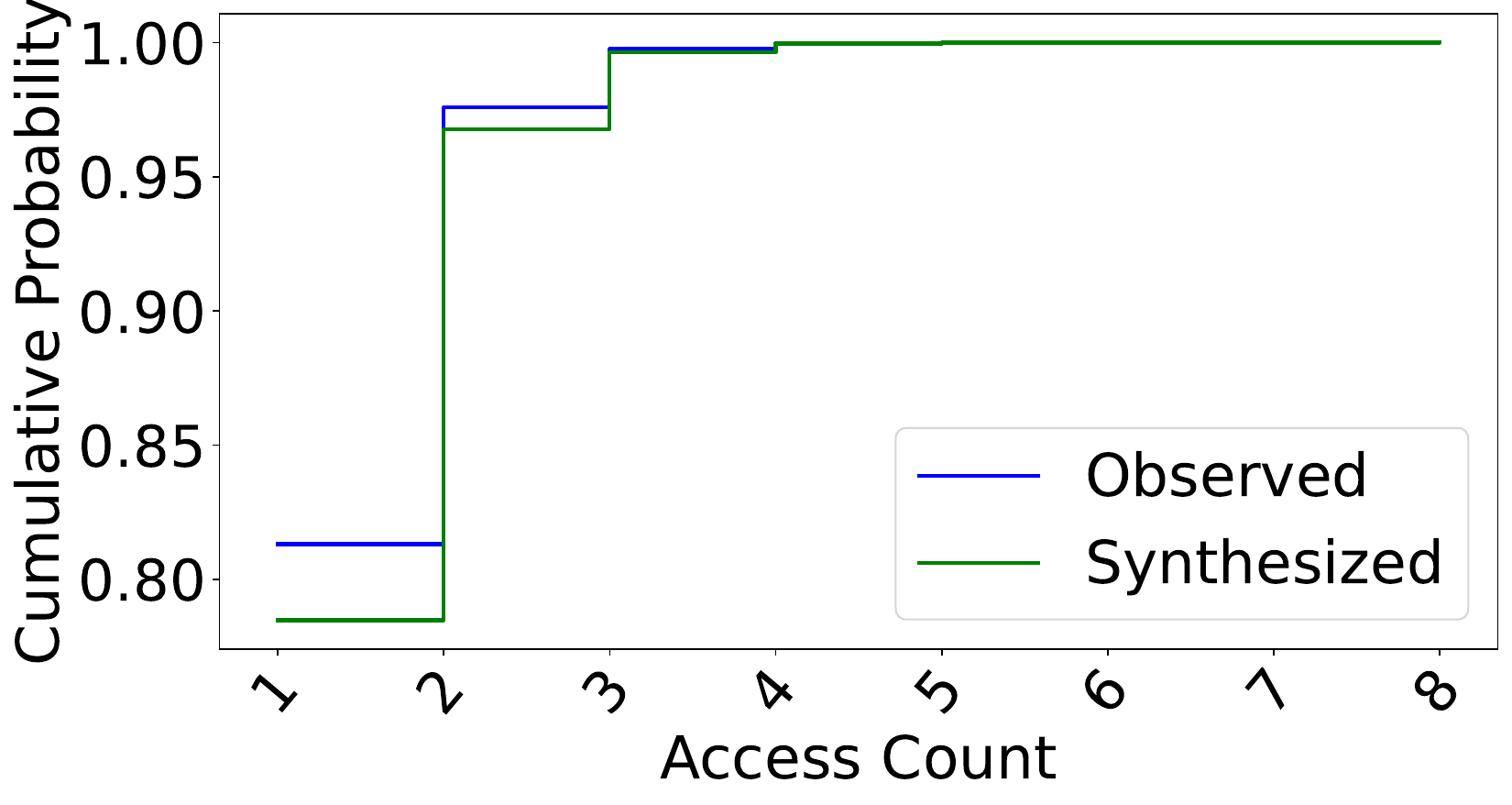}
    \subcaption{Access count distribution}
    \label{subfig:access_count}
  \end{minipage}
  \hfill
  \begin{minipage}[b]{0.245\textwidth}
    \centering
    \includegraphics[width=\textwidth]{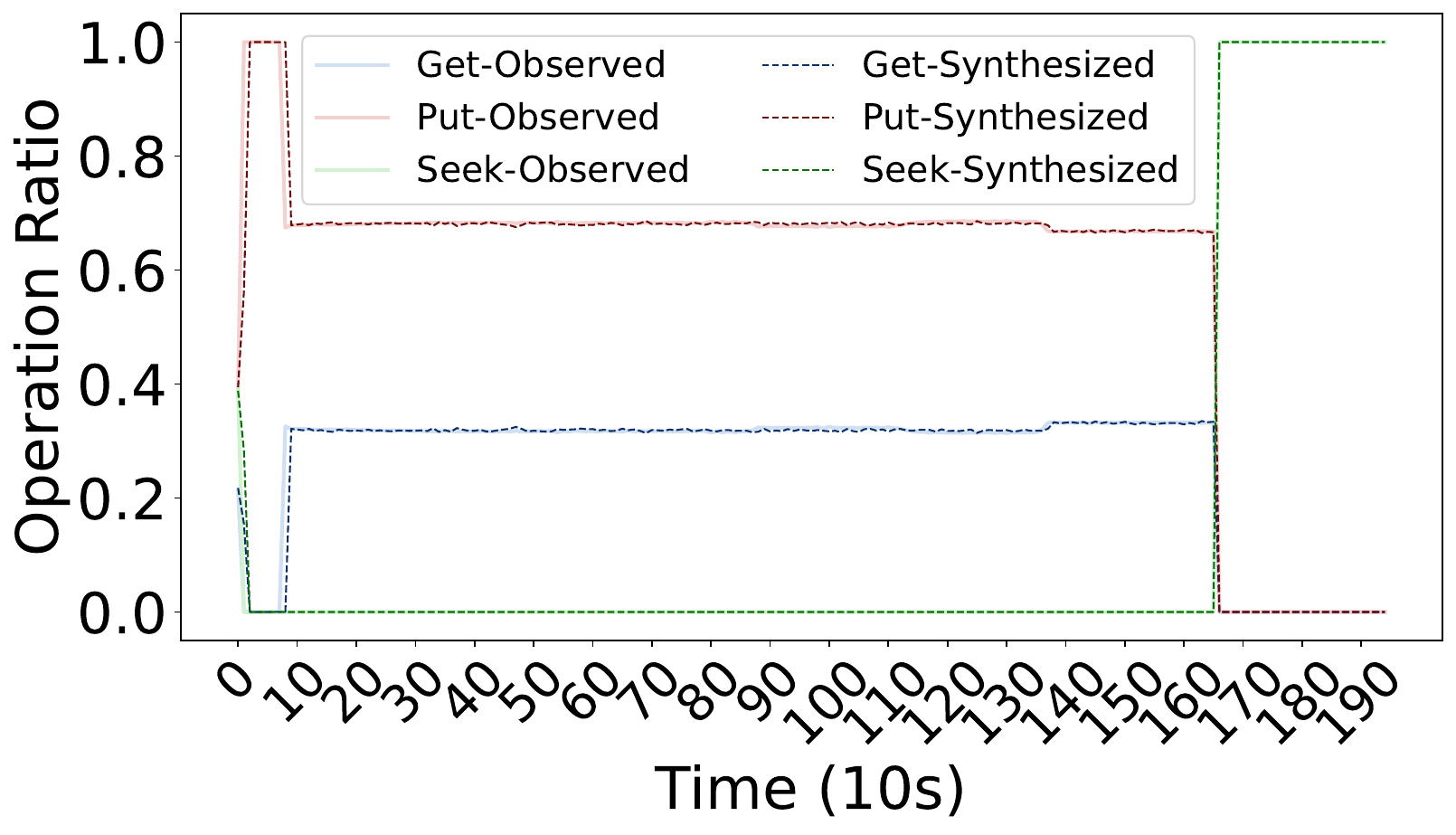}
    \subcaption{Operations ratios}
    \label{subfig:ratio}
  \end{minipage}
  \hfill
  \caption{Comparisons of real-world workloads and synthetically generated workloads in Characterizing and Modeling.}
  \label{fig:characterizing}
\end{figure*}

\begin{table*}[]
    \footnotesize
    \centering

    \caption{Characterizing and Modeling}
    \label{tab:my-table}
    \begin{tabular}{lllllll}
        \hline
        \multicolumn{1}{c}{\textbf{Test}} & \multicolumn{1}{c}{\textbf{}} & \multicolumn{1}{c}{\textbf{\begin{tabular}[c]{@{}c@{}}Key-Size\\      Distribution\end{tabular}}} & \multicolumn{1}{c}{\textbf{\begin{tabular}[c]{@{}c@{}}Value-Size\\      Distribution\end{tabular}}} & \multicolumn{1}{c}{\textbf{\begin{tabular}[c]{@{}c@{}}Key-Access (KA)\\      Distribution\end{tabular}}} & \multicolumn{1}{c}{\textbf{\begin{tabular}[c]{@{}c@{}}Query\\      Distribution\end{tabular}}} & \multicolumn{1}{c}{\textbf{\begin{tabular}[c]{@{}c@{}}KA Distribution\\      $p$-value\end{tabular}}} \\ \hline
        \multirow{2}{*}{Kvrocks} & Observed & Fixed (16b) & Fixed (128b) & Two-Term-Exp & 0.5W:0.5R & \multirow{2}{*}{0.9718} \\
         & Synthesized & Fixed (16b) & Fixed (128b) & Two-Term-Exp & 0.5W:0.5R &  \\ \hline
        \multirow{2}{*}{MixGraph} & Observed & Fixed (48b) & Pareto & Two-Term-Exp & 0.14W:0.83R:0.03S & \multirow{2}{*}{0.9917} \\
         & Synthesized & Fixed (48b) & Pareto & Two-Term-Exp & 0.14W:0.83R:0.03S &  \\ \hline
        \multirow{2}{*}{YCSB} & Observed & Fixed (16b) & Fixed (100b) & Zipfian & 0.5W:0.5R/0.05W:0.95R/0W:100R  & \multirow{2}{*}{0.9778} \\
         & Synthesized & Fixed (16b) & Fixed (100b) & Zipfian & 0.5W:0.5R/0.05W:0.95R/0W:100R &  \\ \hline
    \end{tabular}

\end{table*}

\vspace{0.3em}
\noindent\textbf{ELMo-Tune-V2 Setup with Upper-Level Applications.}
We perform two modifications to make the two upper-level applications, Kvrocks and NebulaGraph, compatible with ELMo-Tune-V2. First, the source code of each application was modified to allow inputs of the modified RocksDB options file for benchmarking. Second, we collected the query-level traces in the initial run, which were directed to an in-memory file system (tmpfs) to ensure minimal tracing overhead. When running benchmarks for the two applications, the process begins by loading the given options file and generating a trace file during benchmark execution. Note that the tracing function is only enabled once the trace is collected. ELMo-Tune-V2 can analyze the generated trace file, characterize and model the current workload, tune the options file, and periodically replace the original options file. These applications can then dynamically adopt the updated options file to serve the current workload better.

\vspace{0.3em}
\noindent\textbf{Baseline Selection.}
We aim to have diverse baselines throughout the evaluation section to compare ELMo-Tune-V2 against. For characterization, there is a lack of baselines present; we instead perform an evaluation using static tools as our baseline and provide statistical analysis like the coefficient of determination to demonstrate our performance.

We use the default configuration file provided by each application as our primary baseline for our tuning evaluations. Examples include the default db\_bench configurations (referred to as \textbf{RocksDB-Default}), the default configurations of NebulaGraph (referred to as \textbf{NebulaGraph-Default}), and the default configurations of Kvrocks (referred to as \textbf{Kvrocks-Default}). In addition to these baselines, we evaluate the performance of our framework against two additional approaches: \textbf{ELMo-Tune}, a solution that utilizes large language models (LLMs) for tuning, and \textbf{Human-Tuned}, a configuration file manually optimized based on the RocksDB Tuning Guide \cite{rocksdb_tuning_guide}. The optimized configuration file generated by our framework is referred to as \textbf{ELMo-Tune-V2}. We provide these configurations for reference as a part of our GitHub repository.

\subsection{Overall Evaluation}

\noindent\textbf{Upper-layer applications.}
We evaluate ELMo-Tune-V2's end-to-end effectiveness in upper-layer applications using NebulaBench and Kvrocks. We utilize ELMo-Tune-V2's characterizing, self-configuring benchmark, and tuning mechanism for each application to generate the RocksDB option files. Further, to ensure a fair evaluation of performance under an out-of-box setup, we also include the default option configuration as provided by NebulaGraph and Kvrocks, presenting them as \textbf{NebulaGraph-Default} and \textbf{Kvrocks-Default} in the results. The results are presented in Figure \ref{fig:upper-layer-apps} and highlight ELMo-Tune-V2's ability to throughput and reduce tail latencies by 25.5\% and 29.7\% respectively when compared to the \textbf{NebulaGraph-Default} configuration in our NebulaBench test, and 20.6\% and 14.45\% when compared to \textbf{Kvrocks-Default} configuration in our Kvrocks test case. Compared to \textbf{Human-Tuned} and \textbf{ELMo-Tune}, \textbf{ELMo-Tune-V2}, on average, outperforms them by 30.1\% and 27.6\%, respectively.

\vspace{0.3em}
\noindent\textbf{Macro Benchmarks.}
We assess ELMo-Tune-V2's performance in real-world benchmarks using MixGraph and YCSB workloads (A through D). For each benchmark, \textbf{ELMo-Tune-V2} integrates its profiling, characterization, and auto-tuning capabilities to adapt dynamically to varying workload patterns. For MixGraph workload, as shown in Figure \ref{fig:upper-layer-apps}, \textbf{ELMo-Tune-V2} achieves improvements in throughput up to 1349\% and reduces the latency up to 57.9\% compared with other baselines. The results of the YCSB workload, summarized in Figure \ref{fig:macro-benchmarks-ycsb}, reveal that \textbf{ELMo-Tune-V2} achieves significant improvements, with throughput gains of 26.6\% and reductions in p99 latency by 14.11\% over \textbf{ELMo-Tune}, another LLM-based tuning framework, showcasing its adaptability across diverse scenarios. It is important to note that the \textbf{RocksDB-Default} baseline fails to show significant improvements due to the lack of any block cache allocated. Meanwhile, the \textbf{Human-Tuned} baseline, based on the RocksDB tuning guide \cite{rocksdb_tuning_guide}, allocates one-third of system memory to block cache, which allows it to perform better than \textbf{RocksDB-Default}. However, both solutions still fall short of \textbf{ELMo-Tune-V2} with \textbf{ELMo-Tune-V2} outperforming \textbf{RocksDB-Default} by 920\% and \textbf{Human-Tuned} by 40\%.

\vspace{0.3em}
\noindent\textbf{Micro Benchmarks.}
To further understand ELMo-Tune-V2's adaptability to different synthetic tests, we evaluate it with varying read-to-write ratios (0:100, 25:75, 50:50, 75:25, 100:0). This analysis measures how \textbf{ELMo-Tune-V2} adjusts configurations to optimize performance for both read-intensive and write-heavy workloads. As illustrated in Figure \ref{fig:micro-benchmarks}, \textbf{ELMo-Tune-V2} demonstrates consistent performance improvements over baselines. On average, \textbf{ELMo-Tune-V2} has throughput improvements of 256\% p99 latency reductions of 33\% compared to \textbf{RocksDB-Default} configuration, 53\% and 26\% throughput and p99 improvements over \textbf{Human-Tuned} configuration. Further, we observe throughput improvements of up to 103\% and p99 reduction of 14-33\% compared to our selected SOTA baseline, \textbf{ELMo-Tune}. The results underscore ELMo-Tune-V2's versatility in managing workload heterogeneity tuning challenges.

\subsection{Sensitivity and Breakdown Analysis}

\noindent\textbf{Characterizing and Modeling.}
Effective workload characterization is fundamental to precise and proactive tuning. This test measures ELMo-Tune-V2's ability to accurately model workload attributes such as key distribution, value distribution, access patterns, and operation ratios. Results in Figure \ref{fig:characterizing} reveal a close alignment between synthesized and observed workload behaviors. further Table \ref{tab:my-table} expands on these tests, demonstrating a close alignment ($p$-value$>$0.95) between synthesized and observed workload characterization. A high $p$-value in our evaluation suggests that the differences between the synthesized and observed data are not statistically significant.

\begin{figure}
    \centering
    \includegraphics[width=\linewidth]{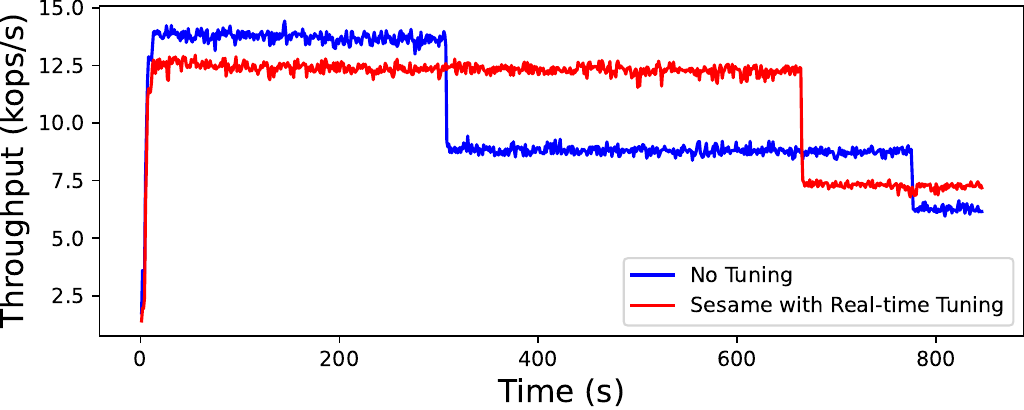}
    \caption{Impact of Real-Time Tuning Modules with Varying Workload (Top: No Tuning, Bottom: Real-time tuning)} 
    \label{fig:tuning-impact}
\end{figure}

\vspace{0.3em}
\noindent\textbf{Impact of Varying Workload for Real-Time Tuning.}
This test examines the impact of the real-time tuning module by synthetically creating a scenario with a shift in the workload pattern. The workload initializes with a throughput-capped write-heavy pattern (100\% Writes), then shifts to a read-heavy pattern (50\% read and 50\& write) after 300 seconds. Finally, at 700 seconds, the workload changes into a more read-heavy workload (85\% reads, 15\% writes). The trace provided to the benchmark in advance omits the shifting pattern of the workload, identifying it as an exclusively write-heavy (100\% writes) workload. With this, we demonstrate the capability of the real-time tuning module, as shown in Figure \ref{fig:tuning-impact} where even without prior workload information, ELMo-Tune-V2 can adjust to the changing workload pattern in real-time. 

\begin{figure}
    \centering
    \includegraphics[width=\linewidth]{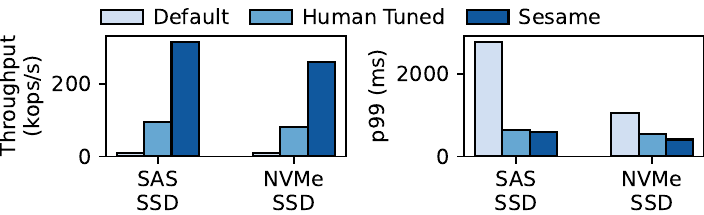}
    \caption{ELMo-Tune-V2 on Different Storage Media} 
    \label{fig:varying-storage}
\end{figure}

\vspace{0.3em}
\noindent\textbf{Varying Storage Devices.}
Storage media diversity introduces unique performance characteristics that influence the performance of RocksDB. This test evaluates ELMo-Tune-V2's adaptability across different storage backends, including a regular SAS SSD and high-performance NVMe SSD, under a real-world workload with mixed read and write patterns (MixGraph). The results, depicted in Figure \ref{fig:varying-storage}, show how ELMo-Tune-V2 dynamically adjusts to storage-specific traits, consistently achieving higher throughput and lower p99 latencies than the baseline. Notably, we exclude HDD due to its considerably lower throughput ($\sim$2000 ops/sec) and higher p99 latency ($\sim$8500 $\mu$s) that would skew the results.


\subsection{Findings and Analysis}

\begin{figure}
    \centering
    \includegraphics[width=\linewidth]{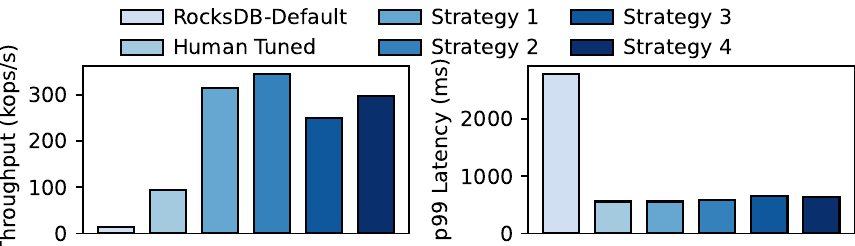}
    \caption{Different Approaches to Tuning Prompts} 
    \label{fig:tuning-prompts}
\end{figure}

\vspace{0.3em}
\noindent\textbf{Different Strategies to Tuning Prompts.}
This test evaluates various strategies for constructing prompts, each representing a distinct strategy for balancing comprehensiveness and focus. The results, visualized in Figure \ref{fig:tuning-prompts}, reveal that every approach is equivalent to performance uplift and latency reduction. We look deeper into the frequency of changes made by the different strategies in Table \ref{table:prompts-table}.

\begin{table}
    \small
    \centering
    \caption{Options Changed Across Iterations for Different Approaches}
        \begin{tabular}{c|c|c|c|c}
        
        & \multicolumn{4}{c}{\textbf{Options Changed}} \\ 
        \textbf{Iteration} & Case 1 & Case 2 & Case 3 & Case 4 \\ \hline
        1 & 11 & 14 & 8 & 6 \\ \hline
        2 & 10 & 4 & 7 & 10 \\ \hline
        3 & 7 & 6 & 13 & 8 \\ 
        \end{tabular}

        \vspace{0.5em}

    \label{table:prompts-table}
\end{table}






\noindent\textbf{Different LLM Models.}
We examine the impact of varying LLM models on ELMo-Tune-V2's effectiveness. Overall, we utilize OpenAI's GPT-4o, 4o-mini, o1-preview, and o1-mini models to perform this test. Preliminary results suggest that ELMo-Tune-V2's characterizing and tuning mechanisms generalize well across models, achieving consistent performance gains owing to the large information corpus integrated into the LLM despite differences in LLM architectures. We conduct these tests using the MixGraph benchmark to offer a real-world test case with a mixed, read-and-write workload and observe a non-significant (less than 5\%) performance deviation between the selected models.



\vspace{0.3em}
\noindent\textbf{Different LLM Parameter Sizes.}
We evaluate the trade-offs between small and large LLM models, selecting the LLaMA series (LLaMA- 3.2-1B, 3.2-3B, 3.1-8B, 3.3-70B models), each with a different amount of parameters it is tuned on to analyze performance difference. Incidentally, the smaller the model size, the more likely it is for the model to give us an incorrect options file - either by not following the format or by simply provisioning incorrect options. In our tests that utilized MixGraph, all models failed to complete the tests and integrate with ELMo-Tune-V2, depicting the necessity of ELMo-Tune-V2 on a large parameter-sized LLM.




\section{Related Work}

\vspace{0.3em}
\noindent\textbf{Workload Characterizing and Modeling.}
Early studies \cite{cao2020characterizing, dong2017optimizing, ycsb} explored workload patterns and focused on providing insights into performance and operational challenges in a corporate setting while also releasing modeled workloads (MixGraph \cite{cao2020characterizing} and YCSB \cite{ycsb}), for the research community to use and learn. However, these workload characterization and modeling studies have relied solely on manual approaches performed by human experts.  
More recently, automated techniques for workload modeling have emerged, leveraging machine learning to generalize across systems. While such methods have shown promise in domains like networking \cite{gan_modeling_network}, they have not been applied to LSM-KVS. 

\vspace{0.3em}
\noindent\textbf{LSM-KVS Tuning and Optimization.}
Several studies have focused on tuning and optimizing LSM-KVS to balance performance and resource utilization. While shown to be effective, existing approaches like ADOC \cite{yu2023adoc}, RTune \cite{RTune}, K2vtune \cite{lee2024k2vtune}, Dremel \cite{Dremel}, CAMAL \cite{yu2024camal}, all focus on a subset of options and workloads using ML-based or heuristic approaches. For instance, ADOC focuses on modifying only two configuration options: the block size and the available resources for background processes. While this method effectively improves performance, it does not assist beyond certain workloads (e.g., write-heavy). In contrast, ELMo-Tune-V2 produces a design that leverages LLM to surpass the subset paradigm set by the current work. Further, our design encompasses an end-to-end solution that includes workload characterization and modeling, along with real-time tuning capabilities not showcased in any current design. 

\vspace{0.3em}
\noindent\textbf{LLM for LSM-KVS Tuning.}
Emerging studies \cite{Thakkar2024can, giannankouris2024lambda, giannakouris2024demonstrating} have explored using LLM for LSM-KVS tuning. For instance, Elmo-Tune \cite{Thakkar2024can} leverages LLMs to generate tuning recommendations by synthesizing knowledge of preset workload patterns, enabling an iterative tuning loop. Meanwhile, $\lambda$-tune \cite{giannankouris2024lambda, giannakouris2024demonstrating} focuses on OLAP applications (e.g., Postgres) to optimize their configurations. Similar goals inspire our approach but differ in two key areas. Unlike both solutions, which rely on pre-trained LLMs for configuration synthesis, our methodology performs characterization and modeling of benchmarks to formulate an end-to-end solution. We diverge from the existing solutions by incorporating real-time tuning, a feature not considered by ELMo-Tune, and focusing on LSM-KVS, which faces very different workloads and more dynamic setups than OLAP applications that $\lambda$-Tune focuses upon.

\section{Conclusion and Future Work}

This paper presents ELMo-Tune-V2, a novel end-to-end framework that performs workload characterizing, modeling, and tuning for LSM-KVS. ELMo-Tune-V2 automates these tasks that are predominantly performed manually, necessitates considerable expertise in the deployment, expected workload, and LSM-KVS, and achieves runtime auto-tuning. This opens new opportunities to characterize, model, and tune other data systems with LLMs. In our future work, we will explore more specific non-obvious configuration parameters by implementing Retrieval-Augmented Generation and fine-tuning the model for specific ecosystem usage.






\bibliographystyle{plainurl}
\bibliography{gpt-references, zhichao-references}

\end{document}